\documentclass{PHYEAUTH}
\usepackage{graphicx}
\usepackage{amsmath}
\usepackage[psamsfonts]{amssymb}
\begin{document}
\begin{frontmatter}
\title{
Metal insulator transition in modulated quantum Hall systems
}
\author{M. Koshino\thanksref{thank1}} and
\author{T. Ando}
\address{
Department of Physics,
Tokyo Institute of Technology
2-12-1 Ookayama, Meguro-ku, Tokyo 152-8551, Japan}
\thanks[thank1]{Corresponding author.\\
\quad E-mail: koshino@stat.phys.titech.ac.jp}
%
\begin{abstract}

The quantum Hall effect is studied numerically in modulated
two-dimensional electron systems in the presence of disorder.  Based on
the scaling property of the Hall conductivity as well as the
localization length, the critical energies where the states are extended
are identified.
We find that the critical energies, which are distributed to each of the
subbands, combine into one when the disorder becomes strong, in the
way depending on the symmetry of the disorder and/or the periodic
potential.
\end{abstract}
%
\end{frontmatter}
%
\section{Introduction}
%
The localization due to the disorder
plays a crucial role in the integer quantum Hall effect.
In a two-dimensional (2D) electron system in strong magnetic fields,
the weak disorder makes almost all the states localized,
and the Hall current is carried only by the extended states
left at the center of the Landau band.
The Hall conductivity is exactly quantized when 
the Fermi energy lies on the localized region.
\par
%
When the 2D electron system is subjected to a two-dimensionally
modulated potential,
the energy spectrum has a recursive gap structure
called the Hofstadter butterfly in each of the
Landau levels \cite{Hofs}.
The system exhibits the quantum Hall effect when
the Fermi energy is in each of those gaps,
where the intricate gap structure 
leads to a nontrivial sequence of the quantized 
Hall conductivity \cite{TKNN}.
The nonmonotonic behavior of the Hall conductivity peculiar 
to this system was experimentally 
observed in lateral superlattices patterned on
GaAs/AlGaAs heterostructures \cite{Albr}.
\par
%
It is an intriguing problem how the Hall conductivity is quantized when the disorder is introduced to the Hofstadter butterfly.
The localization problems in the disordered butterfly system have been studied by several authors.
A finite-size scaling 
analysis was performed and the critical exponent was estimated at the center
of the Landau level \cite{Huck}.  
A qualitative discussion on the evolution of the extended states in
the Hofstadter butterfly as a function of the
disorder for several flux states was given \cite{Tan,Yang99}.
\par
%
Recently the Hall conductivity $\sigma_{xy}$ was calculated and the effect of the localization on the quantum Hall effect was studied in the Hofstadter butterfly \cite{Kosh}.
It was shown that $\sigma_{xy}$ becomes independent of
the system size at  $\sigma_{xy} = 1/2$ (in units of $-e^2/h$),
and those fixed points can be identified as the critical energies
in an infinite system.
While the systems in the previous work
have an electron-hole symmetry between positive and negative energies,
we study here the case without this symmetry
to see whether the fixed points are still on $\sigma_{xy} = 1/2$,
and how the asymmetry 
affects the evolution of the critical energies as a function
of disorder.
The electron-hole symmetry occurs 
when both the periodic and the disorder
potentials are symmetric with respect to zero energy.
We first consider a system 
with an asymmetric disorder potential
containing only positive scatters, 
while the periodic potential is left symmetric.
Second, we recover the symmetry for the disorder, but
make the periodic potential asymmetric as expected
in the presence of the electron screening.
\par
%
\section{Formulation}
%
We consider a two-dimensional system in a strong magnetic field
with a periodic potential $V_p$ and a disorder potential $V_d$,
%
$$
H = \frac{1}{2m}(\Vec{p} + e\Vec{A})^2 + V_p + V_d .
$$
%
The band structure is characterized by a parameter $\phi=Ba^2/(h/e)$, 
a number of magnetic flux quanta penetrating unit cell \cite{Hofs}.
We assume that $V_p$ has a square form
$$
V_p = V \Big( \cos\frac{2\pi}{a}x \!+\! \cos\frac{2\pi}{a}y \Big) 
+ V' \Big( \cos\frac{4\pi}{a}x \!+\! \cos\frac{4\pi}{a}y \Big) ,
$$
where $V'$ represents a double period component 
breaking the electron-hole symmetry 
between positive and negative energies.
The disorder potential is taken as 
randomly distributed delta-potentials $\pm \upsilon_0$,
where the amounts of the positive and negative scatterers 
are given by $N_+$ and $N_-$, respectively.
The energy scale for the disorder 
is given by $\Gamma = 4n_i \upsilon_0^2/(2\pi l^2)$,
where $l$ is the magnetic length and $n_i$ is the number of the
scatterers in a unit area.
We consider only the lowest Landau level, assuming that 
the magnetic field is strong enough
and the mixing of the Landau levels is neglected.
\par
%
We calculate the Hall conductivity using the Kubo formula
for zero temperature,
%
$$
\sigma_{xy} \!=\! \frac{\hbar e^2}{iL^2} \!\!\!
\sum_{\epsilon_{\alpha} < E_F} \sum_{\beta \neq \alpha}
\!\! \frac{\langle \alpha | v_x | \beta \rangle 
\langle \beta | v_y | \alpha \rangle 
\!-\! \langle \alpha | v_y | \beta \rangle \langle \beta | v_x | \alpha \rangle}
{(\epsilon_{\alpha} - \epsilon_{\beta})^2},
\label{Kubo}
$$
%
where $\epsilon_{\alpha}$ is the energy of the eigenstate $|\alpha\rangle$, $E_F$ the Fermi energy, $v_i$ the velocity operator, and $L$ is the system size.
\par
%
\begin{figure}
\leavevmode\includegraphics[width=60mm]{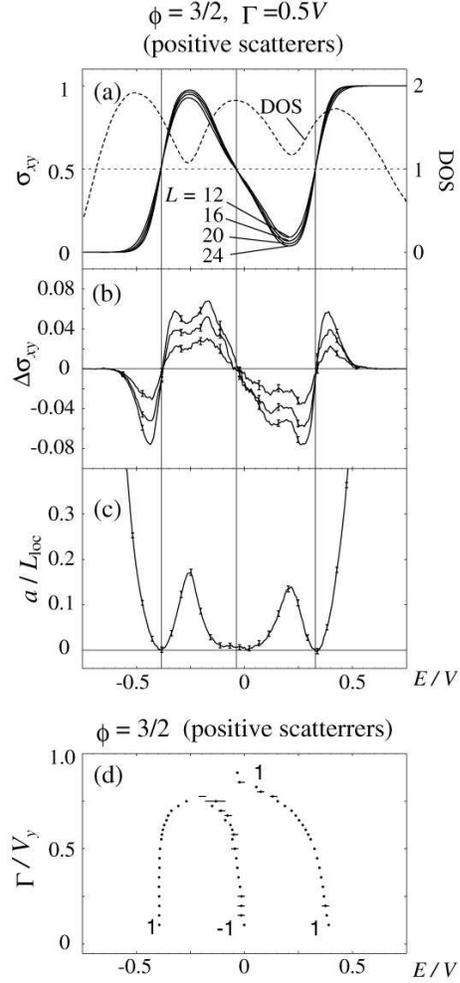}
\caption{
(a) Hall conductivity $\sigma_{xy}$ (in units of $-e^2/h$)
calculated for disordered systems 
containing positive scatterers only,
with $\Gamma/V = 0.5$, $\phi = 3/2$ 
and $L/a=12$, 16, 20, 24.
Dashed line represents the density of states in units of $1/(V a^2)$.
(b) Relative values of $\sigma_{xy}$ measured from
the smallest ($L/a=12$) sample.
(c) Inverse localization length estimated in a Thouless number method.
Vertical lines penetrating the panels represents the energies of
$\sigma_{xy} = 1/2$.
(d) Trajectories of the critical energies as a function of $\Gamma$.
Numbers represent the corresponding Hall conductivities.
Horizontal bars show the energy regions where
the error bar of the Hall conductivity reaches $\sigma_{xy} = 1/2$.
}
\label{fig_iso32pos}
\end{figure}
%
\begin{figure}
\leavevmode\includegraphics[width=60mm]{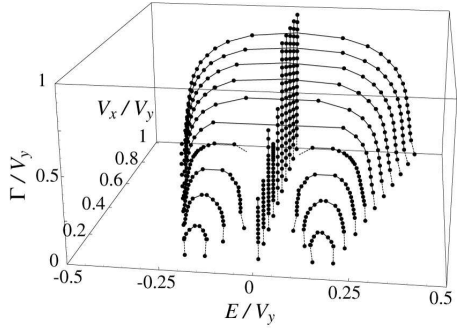}
\caption{
Trajectories of the energies of $\sigma_{xy} = 1/2$ 
as a function of $\Gamma$
and the anisotropy $V_x/V_y$, for a flux
$\phi=3/2$ \cite{Kosh}. The points at $\Gamma = 0$ show the energies 
of $\sigma_{xy} = 1/2$ in clean systems.
At $V_x/V_y = 0.4$, we omit the points $0 < E/V_y < 0.2$
because it is numerically 
difficult to resolve the points around $E=0$.
}
\label{fig_phasebox}
\end{figure}
%
\begin{figure}
\leavevmode\includegraphics[width=70mm]{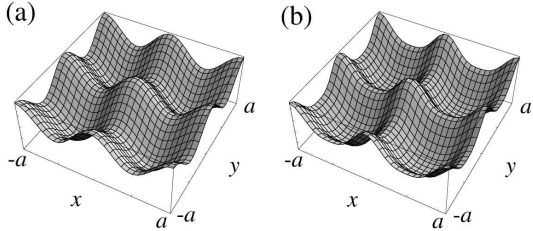}
\caption{
Plot of the periodic potential $V_p(x,y)$ 
with (a) $V'/V = 0$ and (b) 0.2.
}
\label{fig_vp}
\end{figure}
%
\begin{figure}

\leavevmode\includegraphics[width=60mm]{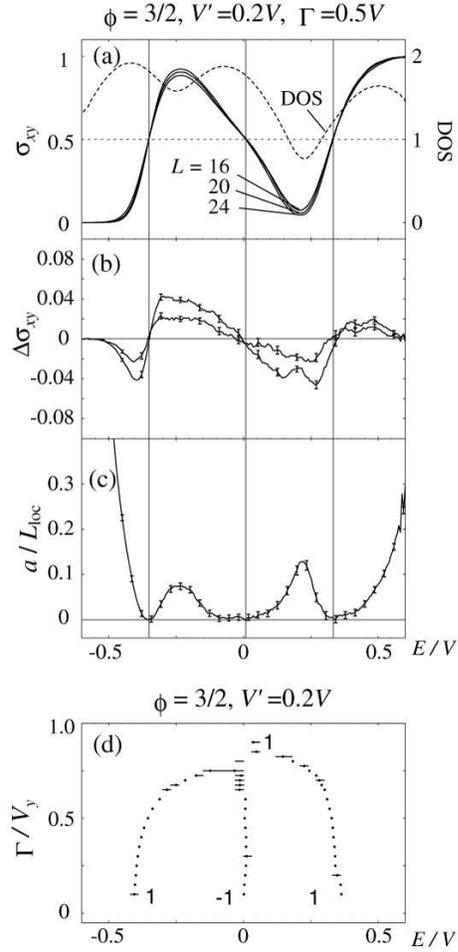}
\caption{
Plots similar to Fig. \ref{fig_iso32pos}
calculated for a system with an asymmetric modulation
$V' = 0.2 V$, $\Gamma/V = 0.5$, $\phi = 3/2$ 
and $L/a=16$, 20, 24.
}
\label{fig_iso32anti}
\end{figure}
%
\section{Results}
%
We first consider a system with symmetric potential $V'=0$
and disorder potential containing only positive scatterers,
where the latter breaks the electron-hole symmetry.
We assume a magnetic flux $\phi = 3/2$, where the lowest Landau level splits into three subbands in the absence of disorder and $\sigma_{xy}$ for each of them becomes $(1,-1,1)$ (in units of $-e^2/h$) \cite{TKNN}.
\par
%
Figure \ref{fig_iso32pos} shows numerical results calculated for the disordered systems with $\Gamma = 0.5 V$.
We show in the panel (a) $\sigma_{xy}$ for several system sizes with the density of states, (b) the difference in $\sigma_{xy}$ measured from the smallest sample, and (c) the inverse localization length $1/L_{\rm loc}$ estimated by the Thouless number method \cite{Ando83}, where every quantity is averaged over a number of different samples.
We can immediately see that each subband has fixed points in the Hall conductivity at $\sigma_{xy} = 1/2$, and they agree with the energies where localization length diverges within the statistics error.
$\sigma_{xy}$ off the fixed points always approaches 1 in the area $\sigma_{xy} > 1/2$, and 0 in $\sigma_{xy} < 1/2$, leading to the Hall plateau in an infinite system.
Therefore it is natural to identify the positions of the extended states as the points of $\sigma_{xy}=1/2$.
\par
%
When we increase the strength of the disorder,
the subband structure becomes obscure and the 
Hall plateau inside the subband gaps should disappear.
The panel (d) in Fig.\ \ref{fig_iso32pos} shows the trace of
the critical energies as a function of $\Gamma$,
which are identified as the positions of $\sigma_{xy}=1/2$.
We can see that 
three energies with the Hall integer $1,-1,1$ become closer 
as the disorder increased, 
and the lowest and the center are united 
at $\Gamma \approx 0.8 V $ and the upper 
is left with the Hall conductivity $+1$.
\par
%
If we distribute equal amounts of positive and negative scatterers,
three critical energies combine all together at $E=0$
due to the electron-hole symmetry.
In Fig.\ \ref{fig_phasebox}, we show the trace of the critical energies
calculated for the symmetric cases for comparison \cite{Kosh}.
We also show the results for anisotropic modulations $V_p = V_x \cos(2\pi x/a) + V_y \cos(2\pi y/a)$ with $V_x \neq V_y$.
When we change the modulation from 2D ($V_x/V_y=1$) toward
1D ($V_x/V_y=0$) with fixed $\Gamma$, 
the critical energy for the center subband
branches into three at $V_x/V_y \approx 0.4$.
Then we have five critical energies for $V_x/V_y<0.4$, even though there are only three subbands \cite{Kosh}.
\par
%
The electron-hole symmetry breaks as well when the modulation potential
itself lacks the symmetry with respect to zero energy.
We consider here a system which has a double period component 
$V' = 0.2V$ in the periodic potential.
The amounts of positive and negative scatterers are assumed to be equal.
The periodic potential is shown in Fig.\ \ref{fig_vp}, where the potential becomes flat at bottom and sharp on top.
The potential in a real system actually has this feature due to the screening by the electrons, even when the external modulation is symmetric \cite{screening}.
\par
%
The numerical results are presented in Fig.\ \ref{fig_iso32anti}.
The Hall conductivity has fixed points
at $\sigma_{xy}=1/2$ in the middle and the lower subband,
while it is hard to decide one
for the upper subband due to a large statistic error.
Possibly this difficulty comes from the low
density of states around this region.
The trajectories of the critical energies, the panel (d), shows that
the center branch is slightly closer to the upper than to the lower
in the region $\Gamma \lesssim 0.5$,
even though the upper subband is a little 
apart from the other two as seen in the density of states (a).
When $\Gamma$ becomes larger, however, 
it is likely that the center branch combines with the lower, 
similarly to the previous case. 
In both cases we have considered, the pair annihilation occurs to
the pair on the side of higher density of states.
\par
%
\section{Summary and Conclusion}
%
To summarize, we calculated the Hall conductivity and the localization
length in the systems without the electron-hole symmetry.  We considered
two different cases, where we introduce the asymmetry to the disorder
potential or to the periodic modulation.  In both cases, we found fixed
points in the Hall conductivity at $\sigma_{xy} = 1/2$, and that those
points coincide with the divergence of the localization length within
the statistics error.  The evolution of the critical energies also
becomes asymmetric in each case in such a way that two branches on the
side with a higher density of states combine together and one remains
intact. We note that, to detect the asymmetry proposed here,
the magnetic field should be strong enough that 
Landau levels are well isolated,
since the asymmetry is also caused by
the coupling among different Landau levels.
We expect that detecting the asymmetry in $\sigma_{xy}$ 
provides a possible probe for
the electron screening in modulated quantum Hall systems.

\par
%
\section*{Acknowledgments}
%
This work has been supported in part by a 21st Century COE Program at
Tokyo Tech \lq\lq Nanometer-Scale Quantum Physics'' and by Grants-in-Aid for Scientific Research and for COE (12CE2004 \lq\lq Control of Electrons by Quantum Dot Structures
and Its Application to Advanced Electronics''),
and Scientific Reseach from the Ministry of
Education, Science and Culture, Japan.  Numerical calculations were
performed in part using the facilities of the Supercomputer Center,
Institute for Solid State Physics, University of Tokyo.
\par
%

%
\end{document}